# Direct Laser Interference Patterning of Functional Metal Surfaces: From Written Geometry to Functional Interfaces

P. Hauschwitz[1*]

*[1]HiLASE Centre, Institute of Physics of the Czech Academy of Sciences, Za Radnici 828, Dolni Brezany 252 41, Czech Republic.*

**Corresponding author: petr.hauschwitz@hilase.cz*

## Abstract

Direct laser interference patterning (DLIP) can generate periodic micro- and nanoscale surface structures with increasing precision and throughput. Yet similar periodic geometries frequently produce fundamentally different functional responses, including reduced friction, modified wetting and ice adhesion, altered bacterial retention or killing, cell guidance, optical effects, and improved electrochemical or photovoltaic performance. These observations cannot be explained by geometry alone, indicating that functional behaviour emerges through a more complex sequence than morphology–performance correlations imply. This review examines that sequence using DLIP as a model system in which the optically prescribed geometry can be separated from the interface that ultimately develops during processing. The written period is distinguished from the realised functional interface, defined by relief depth, aspect ratio, hierarchical topography, surface chemistry, ageing and process history. Functional response is interpreted as a two-stage process in which geometry creates the opportunity for interaction with an external agent, whereas the realised interface determines how that interaction is translated into measurable performance under a specific interfacial state. A curated, metals-centred evidence base spanning tribology, wetting and anti-icing, antibacterial and biomedical surfaces, optics, electrochemistry, energy devices and manufacturing demonstrates that period–depth coordinates alone rarely predict function across application domains. Geometry remains highly transferable in optical systems and controlled mechanical contacts, whereas state-dependent, biological and device-level functions additionally depend on surface state, operating conditions and system architecture. Interpreting these domains through a common causal framework reconciles apparently conflicting observations and identifies which variables remain transferable across applications and which are inherently context-dependent. The resulting perspective provides a common logic for designing and interpreting functional DLIP surfaces: identify the interacting agent, distinguish written geometry from the realised interface, isolate the variables governing interfacial coupling, and evaluate function using direct, mechanism-specific endpoints. Predictive surface engineering and industrial translation therefore depend not only on writing periodic structures reproducibly, but on controlling and verifying the complete realised functional interface.

## 1. Introduction - From Periodic Textures To Predictive Design

Similar periodic surface geometries can produce fundamentally different functional responses. A micrometre-scale relief may reduce friction, delay icing, repel water, retain bacteria, enhance Cu-mediated bacterial killing, guide mammalian cells, trap light or generate structural colour [1–6]. None of these observations is unexpected when considered within its own application domain. Taken together, however, they expose a broader question. If comparable surface geometries lead to different outcomes, which properties of the surface actually determine function?

The difficulty lies in the fact that function is usually discussed through application-specific performance metrics. Tribology reports friction and wear [1], antibacterial studies quantify retention or viable count [3,7], wetting and anti-icing studies measure contact angle or ice adhesion [2,8], while optical surfaces are evaluated through diffraction efficiency [6]. Photovoltaic studies add device-level metrics such as power-conversion efficiency. These endpoints describe different physical, chemical and biological phenomena and cannot be compared on a common scale. As a consequence, the growing body of functional-surface literature provides an increasingly rich catalogue of successful applications but only limited guidance for designing the next one. This limitation reflects a more fundamental distinction between geometry and function. A periodic texture does not interact directly with friction, bacteria or photovoltaic efficiency. It interacts with an external agent: light, a lubricant film, a liquid droplet, a bacterial cell, a mammalian cell or a transport pathway. The measured function emerges only after that interaction has been established. Understanding functional surfaces therefore requires identifying not only the geometry that is written, but also the interface through which that geometry is read.

Direct laser interference patterning (DLIP) provides an unusually clear system in which to examine this problem. Unlike many surface-engineering approaches, DLIP separates the geometry prescribed by the optical interference field from the interface that ultimately develops during processing. The lateral period is determined directly by the interference geometry [9,10], whereas the realised surface emerges through beam delivery, pulse regime, fluence, accumulated dose and material response, which together determine relief depth, aspect ratio, hierarchical topography and surface chemistry [11–14]. The written geometry and the realised interface therefore represent distinct stages of surface formation, each governed by different variables.

This distinction has become increasingly important as the field has matured. Two decades of DLIP research have established reliable processing strategies and demonstrated applications spanning tribology, wetting and anti-icing, antibacterial and biomedical interfaces, electrochemistry, optics and energy devices [15–19]. Reviews and development perspectives have separately documented process evolution and materials-specific advances [9,10]. The remaining challenge is no longer demonstrating that periodic textures can generate functional behaviour. It is determining which aspects of the realised interface retain mechanistic meaning across these otherwise unrelated functional domains.

The central proposition of this review is that functional response develops through two consecutive stages. The written relief first creates the possibility for coupling with an interacting agent. The realised interface then determines how that interaction is translated into measurable performance. Geometry therefore defines the opportunity for interaction, whereas the interfacial state determines its consequence.

Throughout this review, variables within the realised interface are considered mechanistically relevant only when their independent perturbation modifies the geometry–function relationship while the realised geometry is held constant. Such variables are referred to operationally as gates. The term therefore does not describe every experimental boundary condition, but only those interfacial states, operating regimes or system architectures that demonstrably alter how geometry is translated into function. Variables that have not yet satisfied this criterion remain candidate gates or ordinary covariates.

The following sections examine this sequence across the major functional domains currently addressed by DLIP. Rather than comparing application-specific performance records, they ask which variables remain mechanistically transferable, which become domain-specific once the realised interface is considered, and where the available evidence is sufficiently complete to support transferable design rules. Period emerges as the most reproducible and transferable design variable [9,20,21], whereas relief depth, hierarchy, surface state and process history increasingly determine whether a given geometry produces the intended function [8,20,22–25]. The same distinction ultimately defines industrial translation, where reliable manufacturing depends not only on reproducing periodic geometry but on delivering—and verifying—the complete functional interface.

# 2. Scope, Literature Landscape and Evidence Logic

The rapid expansion of direct laser interference patterning (DLIP) has created a problem of interpretation. Similar periods, depths and hierarchical surface morphologies are now reported across tribology, wetting and anti-icing, antibacterial interfaces, biomedical systems, optics, electrochemistry and energy devices, although the mechanisms governing their functional response differ fundamentally. A geometry that promotes fluid retention in a lubricated contact may alter wetting through capillary effects, influence bacterial attachment through local contact conditions, or modify an optical response through diffraction. Period and depth therefore provide a useful geometric description, but they do not by themselves establish why a surface functions or whether a result can be transferred to another application.

This review addresses that gap through a source-linked period–depth–state–function framework. The framework separates the geometry prescribed by the interference field from the surface that is actually produced, the interfacial state established during testing or use, and the resulting functional response. Its purpose is to identify which variables remain transferable across applications whose endpoints cannot be compared directly and to distinguish geometry-driven effects from those controlled primarily by chemistry, ageing, material activity, assay conditions or manufacturing history.

### 2.1 Scope and positioning of the review

The review adopts a deliberately selective, mechanism-oriented scope. The main evidence base centres on metallic systems because they provide the broadest common platform across tribological, wetting and anti-icing, antibacterial, biomedical, optical, electrochemical, energy and manufacturing applications. Studies on non-metallic materials are included where they clarify a mechanism, provide a critical contrast or define the limits of the available evidence. Relevant examples include bacterial adhesion on polymers, polymer wettability, stem-cell responses to topography and previous reviews of DLIP on non-metallic materials [5,7,10,26,27].

Existing reviews address complementary aspects of the field. The closest broad comparator surveys DLIP on non-metallic materials [10], while narrower reviews focus on specific application areas such as bioceramics [28]. Development reviews and conference perspectives document technological progress, process scaling and industrial translation [9,29,30]. Adjacent laser-surface literatures provide domain-specific accounts of LIPSS, laser anti-icing, aluminium texturing, biofilm control and medical-device surfaces [31–35]. These studies establish important application knowledge, although they do not organise the evidence through a common mechanism-based framework that links processing conditions, realised surface state and function across domains.

The present review therefore concentrates on a different question: which variables written or induced by DLIP support transferable mechanistic inference across applications with fundamentally different functional endpoints? The objective requires a selective evidence base. Studies were prioritised when they helped to test the proposed framework, resolve apparently contradictory results, expose limitations in reporting or define the boundaries of current understanding. Broader parameter coverage is retained in the accompanying evidence matrix, allowing the main text to focus on mechanistic interpretation while preserving literature traceability.

Related topics, including laser-textured moulds and photocatalytic surfaces, are considered where they clarify DLIP process control, depth uniformity, manufacturing translation or the distinction between DLIP and adjacent laser-surface-engineering approaches [36–38]. Beyond these functions, they fall outside the scope of the review.

### 2.2 Growth and diversification of the DLIP literature

The scale and thematic development of DLIP research were assessed separately from the selective evidence base used for mechanistic synthesis. This separation is important because publication volume describes the activity of a field, whereas mechanistic inference depends on the quality, comparability and completeness of the underlying experiments.

Figure 1 shows that DLIP has developed from a small, predominantly process-oriented research area into a broad field of functional surface engineering. Publication output increased particularly rapidly after 2015. Although the trailing three-year mean reached its maximum in 2022, recent annual output remains several times higher than during the first two decades of the field. The available data therefore indicate consolidation at a high level of activity rather than a return to the earlier, comparatively limited publication rate.

The expansion has been accompanied by substantial thematic diversification. Process physics and morphology remain the largest component of the literature, while distinct research streams have emerged in optics, tribology, wetting, antibacterial surfaces, biomedical interfaces, energy devices and manufacturing. Several of these areas now produce sustained annual publication activity rather than isolated demonstrations. The literature has consequently become larger and more heterogeneous in its mechanisms, materials, experimental conditions and functional endpoints.

This diversification strengthens the need for a transferable framework. Similar periods and relief geometries are increasingly interpreted through different physical, chemical and biological mechanisms. A growing number of publications therefore does not in itself demonstrate mechanistic maturity. It increases the importance of distinguishing the written geometry, the realised surface state, the conditions under which the interface operates and the function eventually measured.

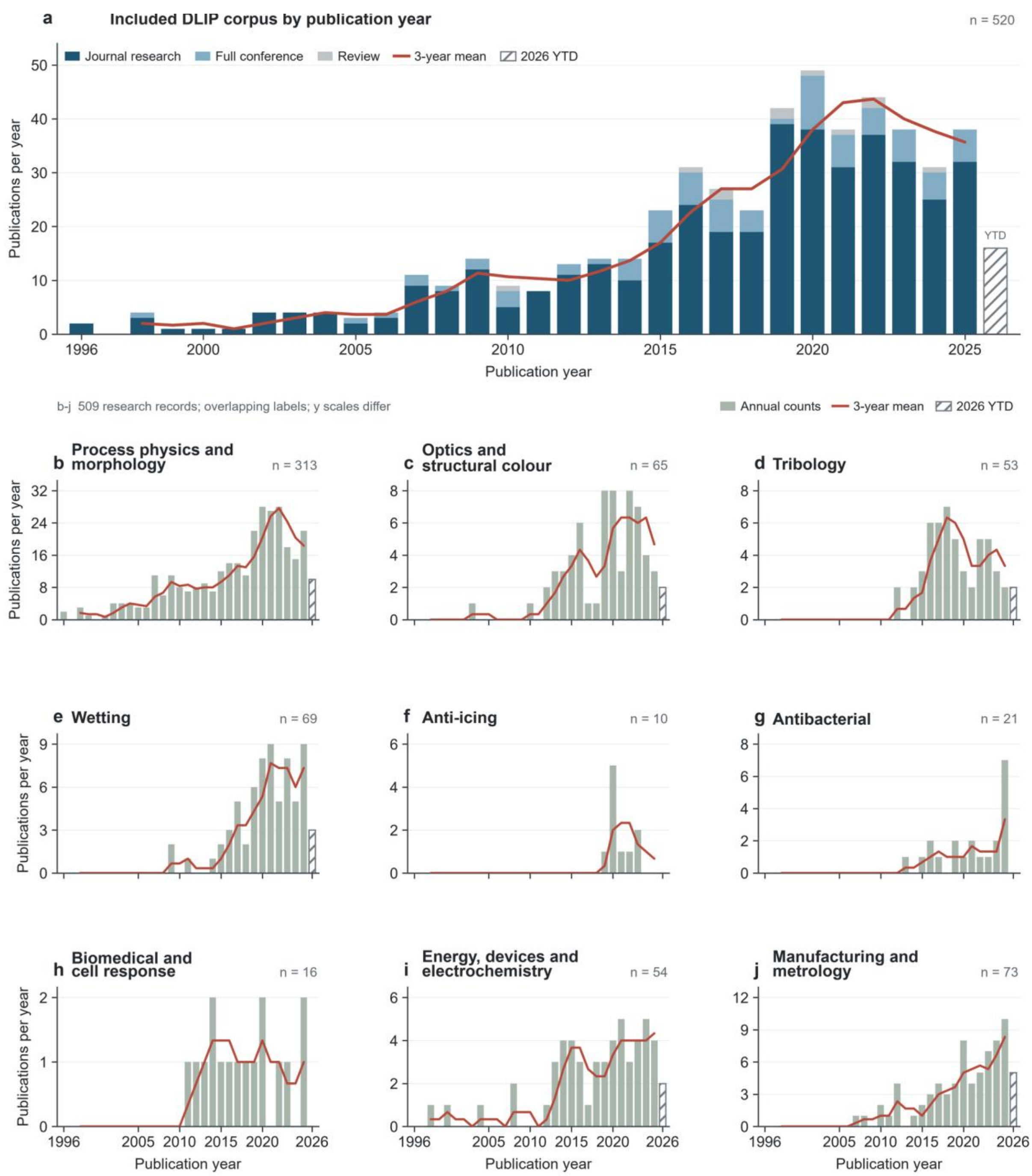

**Figure 1 | a, Annual publication counts within the 520-record bibliometric corpus.** Stacked bars distinguish original journal articles, full conference papers and reviews; the red line shows the trailing three-year mean through 2025. DLIP publication activity expanded markedly after 2015. Although the three-year mean peaked in 2022, recent annual output remains substantially above the levels observed during the first two decades of the field. The hatched 2026 bar reports year-to-date publications through 23 July 2026 and is excluded from the moving average. b–j, Annual publication counts for nine overlapping research themes among 509 original journal and full-conference research records. Topic assignments are non-exclusive, and individual publications may therefore contribute to more than one panel. Reviews are excluded, and 27 research records were not assigned to any of the nine displayed themes. Panel-specific vertical scales preserve trends in lower-volume topics; cumulative n values should therefore be used, rather than bar height, when comparing the overall size of different themes. The trajectories show the development of DLIP from a predominantly process- and morphology-centred field into a broader research area encompassing optical, mechanical, interfacial, biological, energy and manufacturing functions. Publication counts describe field activity and diversification, not evidence strength or mechanistic maturity. Complete search queries, screening decisions, bibliographic metadata and topic assignments are provided in the record-level traceability file.

This expansion has also been accompanied by substantial thematic diversification. Process physics and morphology remain the largest component of the literature, but increasingly distinct research streams have emerged in optics, tribology, wetting, antibacterial surfaces, biomedical interfaces, energy devices and manufacturing. Several of these areas now generate sustained annual publication activity rather than isolated demonstrations. The resulting literature is therefore not only larger, but also more heterogeneous in its mechanisms, experimental conditions and functional endpoints.

### 2.3 Construction of the bibliometric corpus

The bibliometric map was constructed from eight predefined exact-phrase searches of OpenAlex title and abstract metadata. Records were deduplicated and screened individually against predefined inclusion criteria. This procedure retained 516 publications. Four additional in-scope publications were recovered from a predeclared seed collection, producing a final bibliometric corpus of 520 records.

This broad corpus serves a descriptive purpose. It is used to characterise the scale, growth and thematic evolution of DLIP research. Publication counts are not interpreted as measures of evidence quality, scientific importance or mechanistic maturity. The distinction prevents a large application literature from being treated automatically as a mature basis for transferable design rules.

The topic analysis in Figure 1 was performed on 509 original journal articles and full conference papers. Reviews were excluded from the thematic counts, and topic assignments were allowed to overlap because individual studies frequently address several functional or process domains. Twenty-seven research records did not fall within any of the nine displayed themes. Full search strings, screening decisions, bibliographic metadata and topic assignments are retained in the record-level traceability file.

### 2.4 Evidence logic

The mechanistic synthesis requires a more selective standard than the bibliometric analysis. Studies differ substantially in the level of inference they can support. A controlled parameter series linking depth to friction provides a different kind of evidence from a single textured surface shown by microscopy, even when both studies are scientifically rigorous. The evidence classes used here therefore describe the strength and type of design inference supported by each study; they do not rank overall scientific quality.

Four classes are distinguished. Direct experimental evidence links a patterned surface to a measured functional response under a defined assay or operating condition. These studies provide the strongest basis for transferable design inference, particularly when they include parameter series, appropriate references and measurements of the realised surface state.

Model evidence predicts a mechanism, process window or optimum without direct validation of the final function. Such studies help identify plausible causal pathways and candidate design directions, while the predicted relationship remains dependent on subsequent experimental confirmation.

Proxy evidence measures a process- or structure-related quantity, such as focal response, diffraction efficiency, topographical uniformity or another intermediate variable. Proxy measurements can support a proposed mechanism, although they do not establish the final functional outcome.

Morphology-only evidence demonstrates that a texture or hierarchy has been produced without testing its functional consequence. This evidence establishes fabrication feasibility and geometric boundaries but supports only limited functional inference.

Within these classes, greatest weight was given to parameter series, mechanism-bearing contrasts and well-defined functional assays. Such studies reveal how changes in geometry, chemistry, ageing, pulse regime or testing conditions alter the response. Single demonstrations establish feasibility or define boundary cases. Models and proxies support mechanistic interpretation where direct functional data remain unavailable. Direct functional measurements with appropriate controls remain the reference standard for transferable design claims.

This hierarchy also makes contradictions easier to interpret. Two studies reporting similar DLIP periods may yield different functions because their depths, hierarchical features, surface chemistry, ageing state or assay conditions differ. Treating both results as equivalent examples of a period–function relationship would conceal the variables that actually govern the interface. The evidence logic used here is designed to keep those distinctions visible.

### 2.5 Source-linked evidence base and quantitative synthesis

Two literature sets therefore serve different purposes. The 520-record bibliometric corpus describes the breadth and evolution of DLIP research. A smaller source-linked collection supports mechanistic and quantitative inference. Keeping the two sets separate prevents publication volume from being mistaken for evidence strength and allows the quantitative synthesis to remain auditable at the level of individual source values.

The source-linked collection comprises 131 unique DOI records and is not presented as a PRISMA-style systematic review. It was assembled to identify studies capable of supporting source-level interpretation of the relationships between processing conditions, realised surface properties and functional response. Within this collection, 38 records were provisionally classified as direct-function studies.

Source-locked extraction yielded 122 experimental conditions and 160 measurements from 33 studies. The condition-resolved synthesis presented in Figure 6 uses 66 endpoint-compatible observations from 14 studies. Inclusion in this synthesis required the relevant source values, a functionally appropriate endpoint and a reconstructable within-study reference. Records without full-text verification, suitable reference conditions or commensurable functional endpoints remain documented in the screening and traceability tables, but they were not converted into pooled effect estimates.

This staged reduction reflects the difference between reporting a functional claim and supporting a quantitative cross-study comparison. Many studies contribute useful mechanistic or boundary evidence without providing the information needed for normalisation against an appropriate reference. The resulting quantitative subset is therefore smaller than the broader source-linked collection by design.

Most records in the curated evidence base do not appear as individually discussed citations in the main text. Instead, they contribute numerical values to the process-window matrix in Section 4 or to the condition-resolved synthesis in Figure 6. The number of studies named in the narrative is consequently smaller than the total evidence base. This approach preserves readability while maintaining source-level traceability for the underlying quantitative analysis.

Together, the bibliometric corpus, evidence classes and source-linked extraction establish the logic used throughout the review. The broad corpus shows how far DLIP has expanded. The curated evidence base determines which claims can support mechanistic interpretation. The period–depth–state–function framework then provides the structure for asking how processing variables become realised interfacial states and how those states govern functional response across otherwise incomparable applications.

# 3. What DLIP Controls

DLIP begins with a deterministic optical field that defines the written geometry before the material responds. In the symmetric two-beam configuration, the lateral period is prescribed by laser wavelength and half-angle, $\Lambda = \lambda/(2 \sin \theta)$ [9,11]. Three- and four-beam configurations extend the same principle to two-dimensional lattices, with symmetry and accessible periods set by the full beam-vector geometry [6]. Within a nominal period and lattice, polarization, relative phase and beam incidence can further modulate the multi-beam intensity distribution, making field symmetry part of the written optical prescription [39]. This separation makes DLIP useful for disentangling written geometry from the functional interface that ultimately develops.

The realised DLIP surface can be described by four variables. Period is optically prescribed. Depth and aspect ratio emerge through absorption, melt flow, ablation, redeposition and pulse accumulation [11–13]. Hierarchy captures additional structure, including LIPSS, redeposited particles, secondary patterning and post-processing roughness [4,14,40]. Surface state encompasses oxide formation, hydroxylation, hydrocarbon adsorption, contamination, coatings, ageing and biological conditioning [3,8,24,41]. These variables arise from different mechanisms and influence function in different ways.

These variables are not interchangeable descriptors. A 5 μm period combined with a depth of 0.1 μm defines a different tribological interface from the same period with a depth of 0.85 μm [15]. Likewise, a 6 μm DLIP texture on copper can continue to change its wetting behaviour as oxide and carbon chemistry evolve, despite unchanged geometry [8]. Material response also determines whether similar feature sizes produce opposite biological outcomes: a 3 μm, 1 μm deep line pattern on chemically active copper can enhance viable-count reduction, whereas bacteria-scale features on an inert polymer can increase bacterial adhesion [4,7]. Across these examples, the written period establishes the opportunity for interaction, while depth, hierarchy and surface state determine how that opportunity is expressed in the functional assay.

The minimum reportable description of a functional DLIP surface extends beyond period. It includes depth or aspect ratio, pulse regime, accumulated dose, material state, and the post-processing or ageing history that defines the realised interface. Whenever the reported function depends on friction, wetting, ice adhesion, bacterial response, cell behaviour or device performance, assay conditions also become part of the surface description. Omitting variables such as depth removes information that often links fabrication to functional response and manufacturing reproducibility.

This hierarchy distinguishes the optically prescribed interference field from the realised functional surface. DLIP writes a spatial field, whereas functional performance emerges from the physical, chemical and biological interface that develops from it. Maintaining this distinction is essential for interpreting how fabrication variables become functional behaviour (Figure 2).

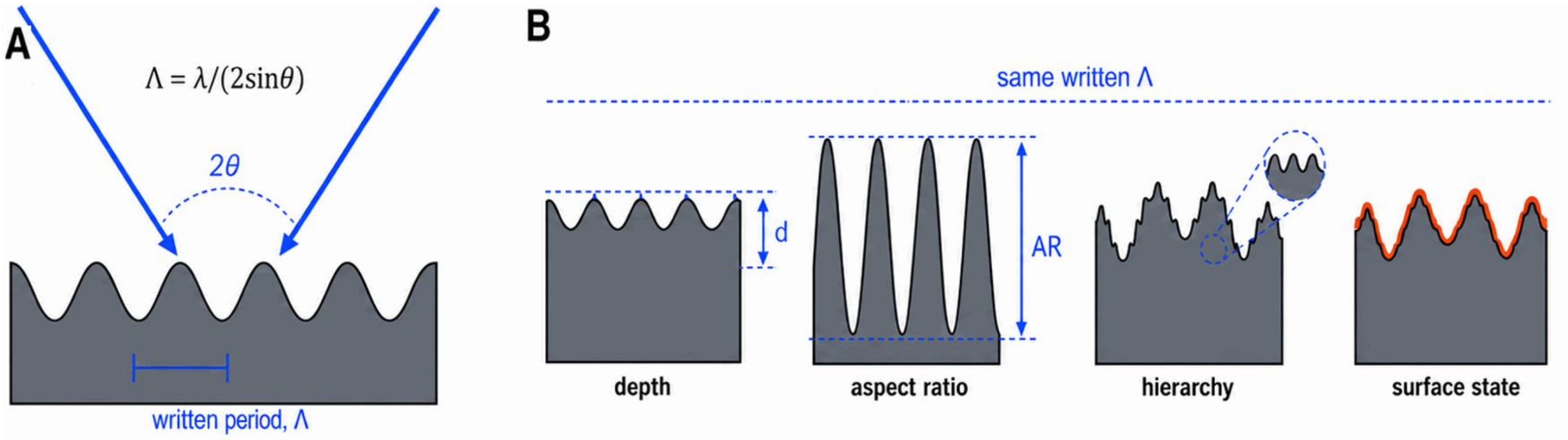

**Figure 2 | Optical prescription and material realisation. a,** In the symmetric two-beam configuration, the optical interference field prescribes the written period, $\Lambda = \lambda/(2 \sin \theta)$, through the beam half-angle θ. **b**, Material response transforms the same written period into a realised interface whose relief depth, aspect ratio, hierarchy and surface state may differ substantially despite an identical prescribed period

**Table 1 | Operational map from realised geometry to a direct functional test.**

| Interacting agent | Coupling variable | Candidate operational gate | Direct endpoint |
|---|---|---|---|
| **Photons** | Period and depth | Illumination and viewing state | Spectral or diffraction response |
| **Lubricant or mechanical contact** | Depth, aspect ratio and orientation | Film and contact regime | Friction and wear |
| **Droplet or ice** | Relief and hierarchy | Chemistry, ageing and environment | Freezing, ice adhesion or de-icing energy |
| **Bacteria** | Contact geometry | Material activity, conditioning and organism | Retention and viable count reported separately |
| **Mammalian cells** | Alignment and access cues | Protein and biochemical context | Alignment, migration and phenotype reported separately |
| **Ions or complete device** | Access and stack geometry | Transport or device state | Electrochemical response or power-conversion efficiency |

A candidate in Table 1 becomes an operational gate only when its independent perturbation modifies the geometry-function relationship at fixed realised geometry; it is distinct from both routine assay conditions and the endpoint used to record function.

# 4. Trends Visible In The Process Window

The DLIP process window is most informative when interpreted as a design map rather than an application taxonomy. Direct experimental evidence shows substantial overlap across functional domains. Structural colour typically uses periods of 1.7-2.6 µm with an optimum depth near 0.3 µm [6]. Lubricated and dry tribology use periods of about 5-18 µm and depths of 0.1-1.13 µm [1,15]. State-dependent functions occupy the same broad range, including 2.6-2.7 µm anti-icing textures, 0.85-3 µm antibacterial textures and 3-17 µm biomedical textures [2–4,25,42]. Period alone therefore does not partition applications into distinct design regimes; similar geometries acquire different roles depending on the interacting agent and interfacial conditions.

Depth is the second dimension of the process window. In optical systems, it tunes diffraction efficiency and colour uniformity once period defines the diffraction geometry [6,43]. In lubricated tribology, performance depends on an aspect-ratio window rather than maximum depth: Bieda et al. reported the strongest response at aspect ratios of about 0.07-0.11, with 25-65% lower friction than unstructured surfaces [15]. In anti-icing, micrometre-scale depth influences wetting-state stability under dynamic conditions [2,42]; in biomedical applications, it governs whether cells bridge, enter or align with features [18,25,44]. Depth is therefore not a universal optimum, but a variable that translates common geometry into application-specific function.

A third dimension is evidence type. Direct functional assays, models and proxies support different inferences even when their reported period-depth coordinates overlap. Missing variables also matter: of 58 source rows in the curated parameter matrix, 42 report a plottable period and depth, 11 lack depth and 5 lack a valid period. These counts describe the audit matrix rather than the entire DLIP literature, but they show why an apparently populated process window can still provide weak mechanistic constraint.

A consistent reporting hierarchy emerges across the DLIP literature. Period is reported most reliably, followed by depth, whereas surface state and assay comparability remain least standardized. This mirrors the technology: period is easiest to prescribe, depth is harder to reproduce, and functional performance is often governed by the least controlled part of the realised interface. The process window therefore identifies the next variable that must be measured or controlled to strengthen mechanistic interpretation and predictive design.

The principal insight is that period alone is a weak predictor across application domains. Stronger predictions require mechanistic context and an explicit account of incomplete evidence. Rather than pointing to a universal optimum, the process window identifies the next variable most likely to improve understanding: depth, surface chemistry, assay design or process metrology (Figure 3).

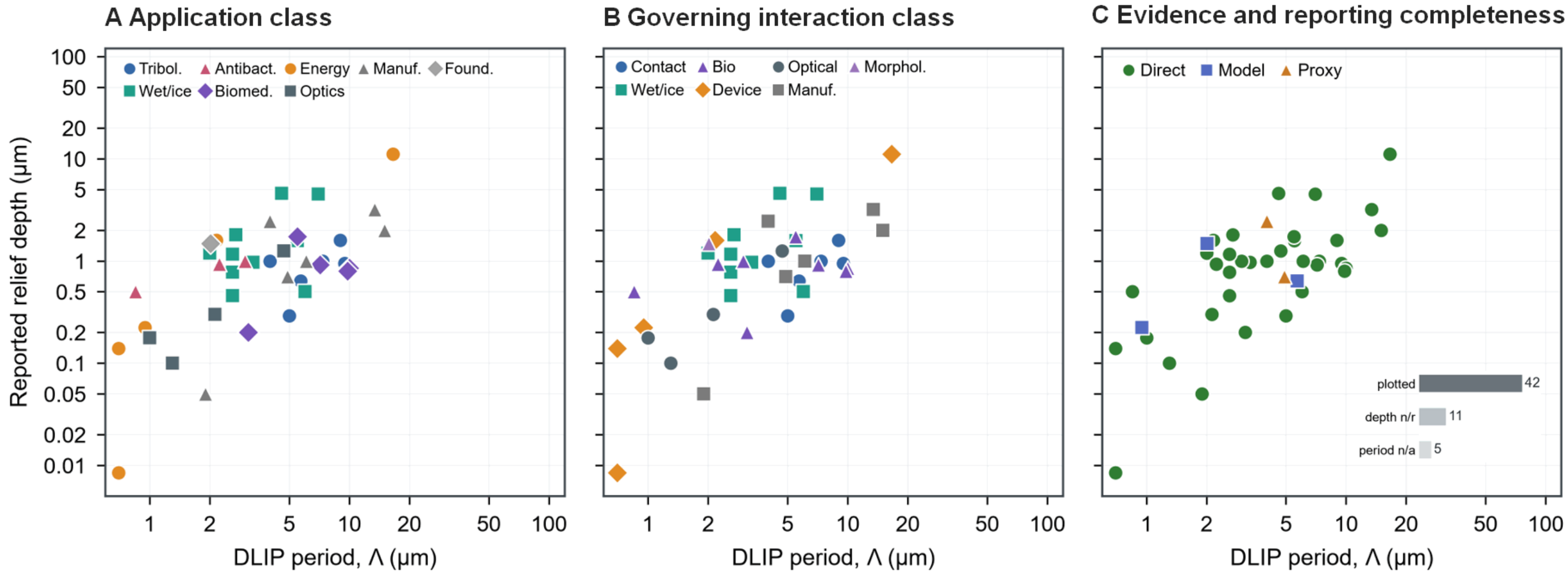

**Figure 3 | The same DLIP process window viewed through application, governing interaction and evidence.** The 42 period–depth coordinates plotted in all three panels overlap extensively, showing that process-window position alone does not identify functional behaviour or determine the mechanistic inference supported by a study. Each symbol represents one source row in the curated DLIP parameter matrix for which both a valid DLIP period and a relief depth could be reconstructed. The coordinates are repeated in **a–c** without aggregation, jitter or displacement; only their classification changes. **a**, Colour and marker shape classify each record by application class: tribology, wetting or anti-icing, antibacterial, biomedical, energy or devices, optics, manufacturing, foundational or mechanistic work, and control or other records. **b**, The identical coordinates are reassigned according to the governing interaction class supported by the source: contact mechanics, wetting or icing state, biological interface, optical coupling, device or transport stack, manufacturing or metrology, or process morphology only. **c**, Marker shape and colour distinguish direct experimental, model and proxy evidence. The inset accounts for all 58 source rows in the curated matrix: 42 with a reconstructable period and depth, 11 without a reported relief depth and 5 without a valid DLIP period. Single reported values are plotted directly. Where a source reports only a bounded period or depth range, the geometric midpoint is shown, while the original bounds are retained in Supplementary Table S1. The extensive coordinate overlap demonstrates that nominally similar period–depth combinations can support different functions and different levels of inference; interpretation additionally requires the governing interaction, the realised interface, the type of evidence and the completeness of reporting.

## 5. Geometry-Rich And Mechanics-Gated Functions

Optical functions provide the clearest demonstration of geometry-dominated behaviour because the interacting agent is a well-defined physical wavelength. Structural colour illustrates this relationship directly. Voisiat et al. used four-beam

DLIP on stainless steel to relate periods of approximately 1.74-2.59 μm to visible diffraction colours and identified an efficient relief depth near 0.3 μm [6]. Dot-matrix holography follows the same principle at smaller length scales: a period of about 1 μm and depths of 0.15-0.21 μm produced first-order diffraction efficiencies of approximately 10%, although discrepancies between scatterometry and AFM highlighted the importance of depth metrology [43]. Optical performance therefore has a clear division of roles: period defines diffraction geometry, whereas depth controls efficiency and colour uniformity.

Lubricated tribology is governed primarily by the vertical geometry of the surface. Relief must be sufficient to modify lubricant storage, contact conditions and debris transport, yet shallow enough to preserve a stable lubricating film. Bieda et al. investigated a 5 μm period on bearing steel with depths of approximately 0.1-0.85 μm and found the greatest friction reduction, about 25-65% relative to unstructured surfaces, at aspect ratios of approximately 0.07-0.11 rather than at maximum depth [15]. Tribological performance is therefore controlled by the interplay between surface geometry and lubrication regime, making aspect ratio more informative than period alone.

Component-scale studies reinforce the same conclusion. Grutzmacher et al. applied a 5.7 μm period with a depth of approximately 0.64 μm to stainless-steel journal-bearing components, shifting the contact toward a more favourable lubrication regime and reducing friction relative to the reference surface [16]. The overall response nevertheless depended on the accompanying dimple geometry, while numerical studies show that load, film formation and contact geometry define the useful operating window [45–47]. Interpreting tribological performance therefore requires friction measurements to be considered alongside period, depth or aspect ratio, and the corresponding contact conditions.

Dry tribology illustrates how a change in contact mechanics alters the functional role of the same geometric variables. Rosenkranz et al. compared line textures with periods of 5, 9 and 18 μm and depths of approximately 0.8-1.13 μm under dry reciprocating sliding, showing that texture orientation and period influenced friction, and that the optimum under dry conditions differed from the lubricated case [1]. Related studies on oil-film lifetime, load-dependent run-in, lubricant migration and contact-area tailoring reach the same conclusion from complementary assays [48–51]. Under lubrication, texture primarily modifies film formation; under dry sliding, its effect is mediated through debris generation, orientation, counterbody interactions and wear evolution.

Optics and tribology establish the clearest mechanistic reference points for interpreting DLIP function. In optical systems, period and viewing geometry dominate the response; in tribology, depth, aspect ratio and contact regime become critical. Geometry is most predictive when the interacting agent and governing interfacial conditions are well defined. As functional behaviour becomes increasingly influenced by chemistry, ageing and biological processes, the same period-depth combination no longer predicts the outcome on its own.

# 6. State-Gated Interfacial Functions

Wetting, anti-icing and antibacterial applications mark the point where geometry alone no longer explains function. The written relief defines contact opportunities, liquid confinement or exposure sites, but the measured response depends on the realised surface state and the conditions under which it is tested. Oxide and carbon chemistry govern wetting, Cassie/Wenzel stability governs anti-icing behaviour, and material activity together with the biological assay determines whether similar topographies promote bacterial retention or viable-count reduction. The wetting literature provides a clear demonstration of this separation. Marie-Loelein et al. applied DLIP to copper using a 6 μm period and a depth of about 0.5 μm, then followed the evolution toward stable superhydrophobicity with contact angles approaching 170°. Their analysis distinguished topography from oxide formation and hydrocarbon adsorption, showing that the same DLIP geometry acquires different wetting behaviour as surface chemistry evolves. Copper wettability therefore cannot be interpreted from geometry alone; storage history and surface chemistry are part of the functional interface [8]. Stainless steel and aluminium studies reach the same conclusion from complementary directions. Aguilar-Morales et al. combined a 5.5 μm DLIP period with multiscale texturing and chemical treatment on stainless steel, achieving a static water contact angle of 152 ± 2° and contact-angle hysteresis of 4 ± 2°; the result depended on chemical modification rather than roughness alone [52]. Milles et al. reported that hierarchical DLIP/DLW textures on aluminium reached contact angles near 160° only after ageing, while freezing delay increased from about 8.7 s to 22 s under the reported conditions [17]. Contact-angle measurements are therefore comparable only when the temporal evolution of the surface is reported with the geometry.

Anti-icing is more demanding because the relevant interfacial state evolves during icing. Static droplet measurements at room temperature do not reproduce impingement, condensation, freezing-front propagation or ice-shear loading. Vercillo et al. evaluated hydrophobized laser-treated Ti6Al4V under aeronautic icing conditions and identified the best DLIP texture at a period of about 2.7 µm and a depth of about 1.8 µm, with ice adhesion of about 11-18 kPa in the reported interfacial-shear assay; other superhydrophobic rough surfaces did not achieve similarly low adhesion and sometimes increased mechanical interlocking [2]. Alamri et al. reached a complementary conclusion using de-icing power: a DLIP surface with a 2.6 µm period and about 0.78 µm depth showed a contact angle of 163 ± 6°, an 8° roll-off angle and a 48-80% reduction in de-icing power under the tested conditions [42]. Anti-icing performance therefore requires functional metrics such as ice adhesion, freezing delay, droplet rebound, roll-off or de-icing power to be reported with morphology, chemistry/post-treatment and icing regime.

Antibacterial DLIP provides one of the clearest sign changes. Valle et al. showed on polymer surfaces that bacteria-scale line and pillar features increased *S. aureus* adhesion, whereas only the lamellar texture reduced adhesion under the tested conditions. The reported periods were 1-5 µm, with feature depths of about 0.47-1.85 µm [7]. Matching a surface feature to bacterial dimensions is therefore insufficient to infer an antibacterial response. Stainless steel and copper show why antibacterial endpoints must remain separate. Peter et al. reported a 0.85 µm period and a depth of about 0.5 µm on 316L stainless steel, reducing two-hour bacterial retention by about 99.8% for *E. coli* and 70-79% for *S. aureus*; this assay quantified early anti-retention rather than killing [3]. Muller et al. used 3 µm DLIP line textures with a depth of about 1 µm on oxygen-free copper, achieving up to a 15-fold reduction in colony-forming units relative to smooth copper and 99.9% killing within about 60-70 min. The dominant mechanism was Cu-mediated bactericidal activity enhanced by contact efficiency rather than purely mechanical disruption [4]. Retention, colony-forming units, live/dead imaging and long-term biofilm formation therefore measure different biological processes. Polymer, stainless steel and copper stop looking contradictory once anti-adhesion, killing and material class are separated.

Across state- and chemistry-dependent functions, geometry acquires meaning only in relation to the intended functional state. Wetting must be interpreted in the context of Cassie/Wenzel state and ageing history, anti-icing in the context of icing regime and functional endpoint, and antibacterial performance in terms of retention, killing or biofilm suppression. Geometry establishes the opportunity for interaction; demonstrating function requires measurements that probe the relevant interfacial mechanism rather than a convenient geometric surrogate.

Biomedical surfaces extend the same logic to adaptive biological systems. Mammalian cells span surface features, form focal adhesions, remodel their cytoskeleton and respond to proteins, ligands, oxide chemistry, substrate stiffness and culture conditions. Abagnale et al. showed that laser-interference nanogrooves with a 650 nm period influenced mesenchymal stem-cell behaviour on polymer substrates. In the same study, however, the lineage-selective 2 and 15 µm microgrooves were produced by reactive-ion etching rather than DLIP and therefore provide mechanistic context rather than direct evidence for DLIP on metallic surfaces [5]. Metal and ceramic studies show that different biological endpoints respond to different parts of the realised interface. Schieber et al. investigated CoCr surfaces with periods of 3, 10, 20 and 32 µm and varying relief depths. Endothelial alignment exceeded 67% only on stronger topographies, whereas platelet reduction occurred even on flat laser-oxidized controls, indicating that alignment and haemocompatibility were governed by different mechanisms [18]. Zwahr et al. studied titanium surfaces with periods of 3, 5, 10 and 17 µm and depths of about 0.4-2.1 µm; the 17 µm texture produced up to about 2.5-fold more osteoblast cells after seven days than the reference surface, whereas smaller periods did not produce the same access and proliferation response [25]. Minguela et al. reported that both 3.03 and 9.92 µm DLIP grooves on Y-TZP induced strong hMSC alignment, while osteogenic differentiation depended primarily on peptide functionalization: alkaline phosphatase increased about 2.5-fold on peptide-treated surfaces, including flat controls [44]. A stainless-steel study combining multi-beam processing with beam shaping reported up to a 496% increase in cell number relative to untreated steel and preferential orientation on combined LIPSS–micropillar surfaces. Because the nano- and microscale features were varied together, the result supports functional feasibility more strongly than an isolated geometric design rule [53]. Biomedical interpretation therefore depends on the endpoint. Alignment, migration, proliferation and differentiation should not be collapsed into a single measure of biocompatibility. DLIP can regulate cell access and guidance, whereas biochemical cues frequently determine subsequent cellular behaviour. Meaningful interpretation requires period, depth, cell-interaction mode, biochemical state and endpoint-specific assays to be considered together.

Across wetting, anti-icing, antibacterial and biomedical systems, the same geometric features can produce fundamentally different outcomes. A water droplet may become mobile or pinned, ice may detach or mechanically interlock, bacteria may be retained or killed, and mammalian cells may align without differentiating. In each case, geometry establishes the opportunity for interaction, whereas the realised interfacial state determines how that interaction is expressed experimentally.

# 7. Mixed Transport And Device Functions

Energy and electrode applications occupy an intermediate position between geometry-dominated and state-dependent functions. In photovoltaic devices, DLIP can modify light scattering, haze and optical path length, but the measured response reflects the complete device stack rather than the textured surface alone. Muller-Meskamp et al. applied periodic DLIP textures to flexible organic solar cells using a 0.7 µm hexagonal pattern and a 4.7 µm line pattern. The hexagonal texture increased power-conversion efficiency by about 21% relative to the flat reference, whereas the line texture improved efficiency by about 5%, with part of the gain arising from fill factor rather than optical absorption alone [54]. Optical enhancement becomes functionally relevant only when it is preserved by the electrical response of the device.

Transparent-electrode work shows the trade-off more directly. Heffner et al. produced FTO textures with a 700 nm period and shallow reliefs of 4-18 nm. Haze increased by about 365-1000%, while perovskite solar-cell efficiency improved from 18.01% to 19.43% under the best-performing condition. At the same time, sheet resistance increased by about 48-63%, so the condition maximizing optical scattering did not maximize device performance [55]. Device efficiency reflects the balance between optical gain and electrical loss rather than either quantity in isolation.

Modelled light-trapping studies occupy a different level of evidence from experimentally validated devices. Soldera et al. explored perovskite textures with periods of about 0.3-3 µm and depths of about 0.1-0.5 µm, predicting an absolute efficiency improvement of roughly three percentage points under the assumptions of their model [56]. Such studies identify promising regions of design space and generate mechanistic hypotheses. Their predictions acquire design significance only after validation within complete devices, where stack architecture, fabrication route and electrical losses are evaluated experimentally.

Energy studies therefore form an evidence chain from optical coupling to device performance. Morphology-only studies establish whether the required texture can be fabricated, modelling identifies candidate optima, and complete device measurements determine whether the predicted optical advantage survives integration into a photovoltaic architecture. Performance can fail at several stages, including optical absorption, charge transport, layer conformity and long-term stability.

Electrode applications add transport and electrochemical surface chemistry. Ranke et al. fabricated high-aspect-ratio periodic structures on nickel with periods of 11 and 25 µm and depths of about 9.4-13.1 µm. The resulting surfaces reduced the hydrogen-evolution overpotential by about 22% at the reported current density, although the available data did not fully disentangle electrochemical surface area, mass transport and intrinsic catalytic activity [19]. Recent studies on nickel electrodes and bubble management reach similar conclusions: surface geometry influences electrolyte access, gas evolution and local transport, while the measured electrochemical response depends on electrolyte composition, oxide state and testing protocol [57–60].

Mixed transport functions are governed by coupled variables rather than geometry alone. Period and depth determine access for light, ions, gas or charge, whereas material state, layer conformity and device architecture determine whether that access improves the final metric. Optical absorption, haze, sheet resistance, overpotential, power-conversion efficiency and catalytic response therefore occupy different levels of functional evidence. Treating them as one generic performance scale hides where gains are created and where they are lost. Figure 4 illustrates this range directly, pairing representative realised surfaces with the corresponding functional observation.

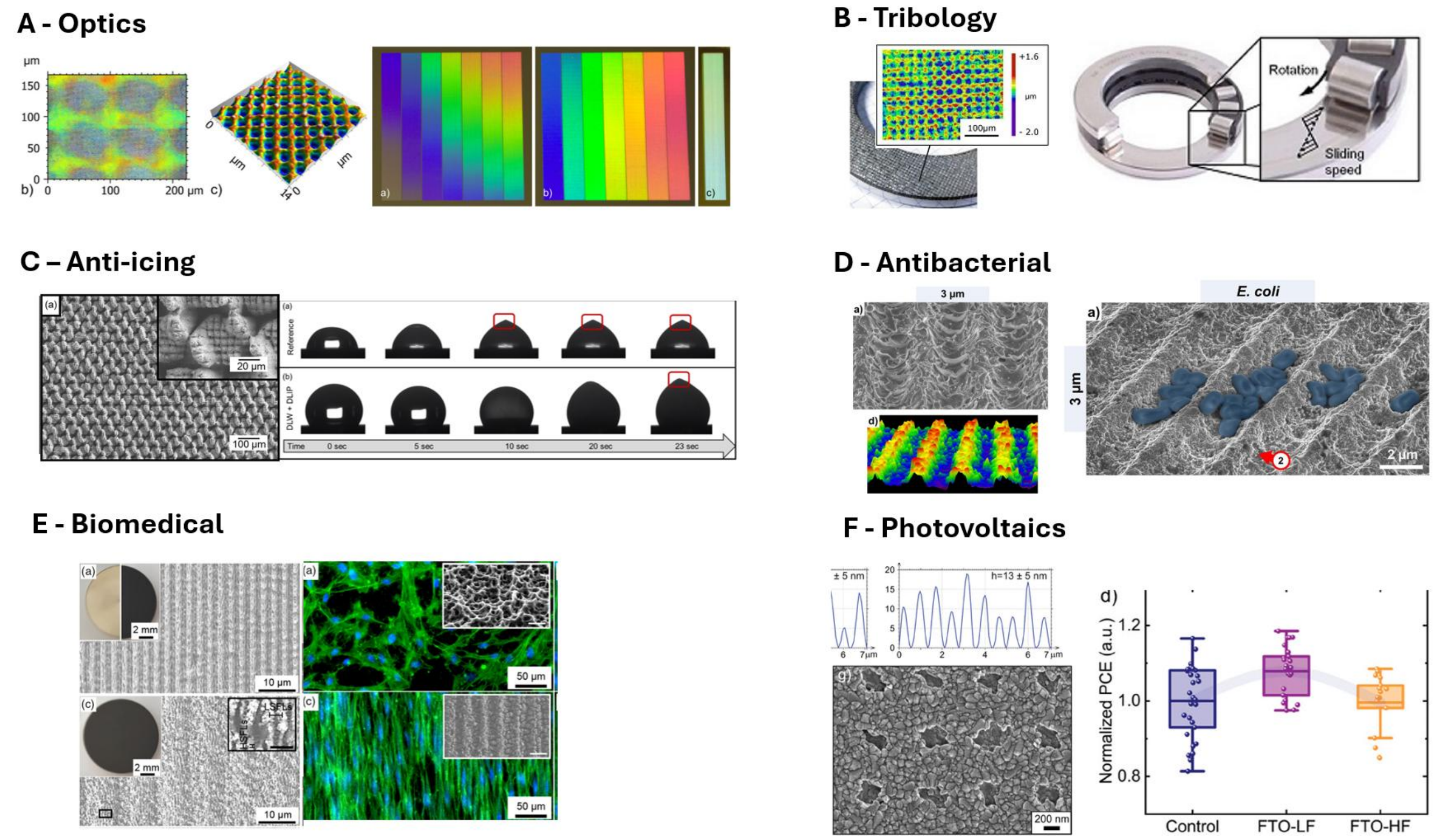

**Figure 4 | Representative realised DLIP interfaces and their application-specific functional expression. a,** Gradient-period structures and corresponding diffraction colours on stainless steel [6]. **b,** Laser-interference-patterned cylinder roller bearings and the associated ZDDP tribofilm response under lubricated contact [61]**. c,** Hierarchical DLW–DLIP aluminium relief and the corresponding droplet-freezing sequence [17]. **d**, A 3 µm USP-DLIP line pattern on copper, its corresponding topographic profile, and SEM observation of E. coli interacting with the patterned surface [62]. e, Periodic titanium implant surfaces and corresponding osteoblast morphology and alignment [25]. f, DLIP-patterned FTO topography and the resulting power-conversion efficiency of perovskite solar cells fabricated on untreated and patterned FTO [55]. Together, the panels illustrate how comparable period–depth descriptions acquire distinct physical meanings through optical, mechanical, liquid, bacterial, cellular and device-level interactions. All source material in panels a–f is adapted under CC BY 4.0. Source panels were cropped, resized, relabelled and combined; the underlying scientific data were not altered.

# 8. Manufacturing Translation

Manufacturing translation extends the same design problem from laboratory surfaces to industrial production: the written geometry must remain coupled to the realised functional interface as processing scales. DLIP is inherently suited to large-area manufacturing because it writes periodic structures in parallel rather than point by point, and large-beam, scanner-based, beam-shaped and roll-to-roll implementations have progressively increased processing capability [9,12,29,63]. Hauschwitz et al. enlarged the interference footprint by shifting the interference plane away from the focal plane, producing an elliptical patterned area with a major diameter above 1 mm. The study reported fabrication-rate limits of 206 $cm^2 min^{-1}$ for single-pulse processing and 0.34 $cm^2 min^{-1}$ for 1000-pulse processing. Morphology varied strongly with material, pulse energy and pulse count, but the reported rate limits were not accompanied by complete depth measurements at the same operating conditions. In a separate 1000-pulse morphology series, sub-1-mJ irradiation produced 2.64 µm grooves approximately 1 µm deep on stainless steel [64]. The same large-beam principle was extended to four-beam ps-DLIP/LIPSS: demonstrated stitching rates reached 10 $cm^2 min^{-1}$ for low-spatial-frequency LIPSS and 68 $cm^2 min^{-1}$ for high-spatial-frequency LIPSS. The higher values of 0.05 and 0.34 $m^2 min^{-1}$ reported in that study were projections for a planned 50 kHz source rather than demonstrated processing rates [65]. Madelung et al. demonstrated scanner-based DLIP on stainless steel with periods of 2.9-12.8 µm, achieving 7.69 $cm^2 min^{-1}$, approximately five times the historical linear-axis benchmark cited by the authors [20]. Zwahr et al. extended this concept to roll-to-roll processing of current collectors, demonstrating about 0.20 $m^2 min^{-1}$ for 15 µm textures; across the study, reported depths were approximately 1-4 µm on aluminium and 1-2 µm on copper [21]. These demonstrations establish industrially relevant processing capability, but also expose the central translation problem: faster writing is useful only when the depth, surface state and functional interface produced at that rate are known.

Scaling the prescribed period is only the first step toward functional manufacturing. The variables that govern performance remain substantially more difficult to reproduce: depth and aspect ratio depend on material response and accumulated dose, while surface state continues to evolve through oxide formation, contamination, coating and ageing. A process that reproduces periodicity while allowing depth distribution or surface state to drift may therefore lose tribological, wetting, optical or biological performance despite unchanged geometry. Manufacturing capability is consequently defined not only by throughput, but by the reproducibility of the realised functional interface. This requirement determines the role of process monitoring. Steege et al. demonstrated photo-acoustic autofocus with a strong correlation between acoustic emission and z-position ($R^2$ about 0.97) for DLIP periods of 2, 4, 6 and 8 µm, providing reliable focus control while leaving the relationship between depth and functional performance to be established separately [23]. Diffractive monitoring, emission monitoring and machine-learning approaches extend this progression toward closed-loop manufacturing [66–68]. Their value ultimately depends on the variable being predicted: recognizing periodic patterns improves inspection, whereas predicting depth distribution, oxide or carbon state, ice adhesion, bacterial retention or friction lifetime enables direct control of the functional interface itself.

Manufacturing maturity therefore has two independent dimensions: processing throughput and control of function-relevant variables. Increasing area rate, from static coupons to scanner-based, on-the-fly and roll-to-roll processing, demonstrates production capability. Reproducible control of period, depth, depth distribution and surface state determines whether functional behaviour can be transferred to industrial scale. A high-throughput process that reports only periodicity therefore provides weaker evidence than one that verifies the variables governing function.

Functional manufacturing in DLIP requires optical, topographic and functional delivery to remain aligned. Optical delivery preserves period, symmetry and field stability. Topographic delivery preserves depth, aspect ratio, hierarchy and uniformity at production rate. Functional delivery preserves surface state, ageing behaviour, durability and in-service performance. Throughput without function-relevant metrology is therefore a measure of production capability, not yet reproducible functional-surface manufacturing. Figure 5 illustrates this trade-off directly, comparing the reported fabrication rates and representative morphologies for large-beam, scanner-based and hatch-optimized DLIP processing.

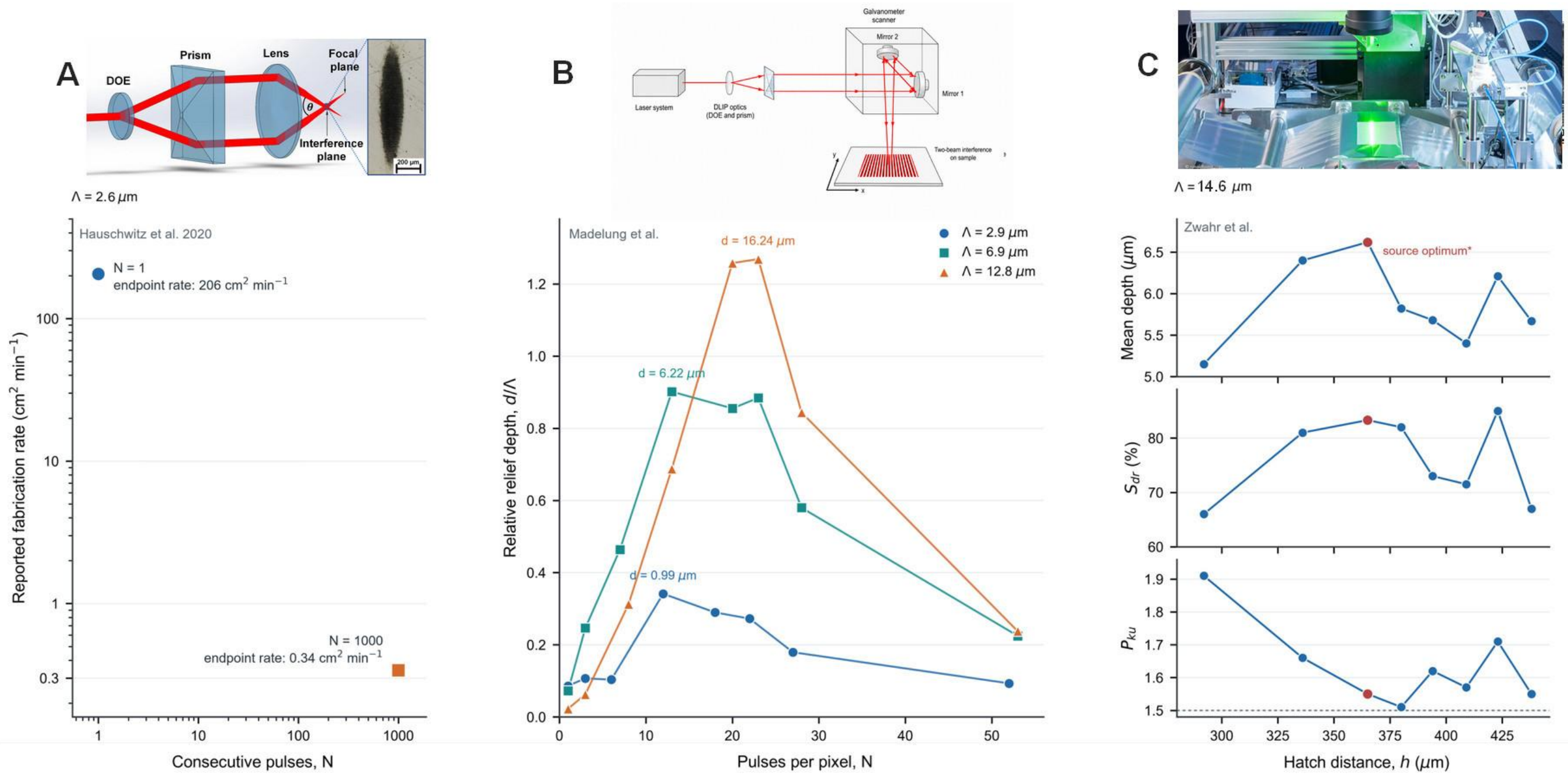

**Figure 5 | Process-rate data and representative morphology.** Scale-up strategies, rate-defining variables and realised morphology in DLIP processing. **a**, Large-beam two-beam DLIP, in which separation of the interference and focal planes enlarges the interference footprint and increases the patterned area per pulse. The upper panels illustrate the optical concept and the calculated interference field for Λ = 2.64 µm. Reported endpoint fabrication rates are shown for single-pulse and 1000-pulse processing. These rate values were not paired with depth measurements at identical operating conditions; in a separate 1000-pulse morphology series, pulse energies below 1 mJ produced approximately 1-µm-deep grooves at the same period [64]. **b**, Scanner-based DLIP of stainless steel at 1.36 J cm⁻². Published relief depths are replotted as relative depth, d/Λ, against pulses per pixel for three spatial periods. Relief development is non-monotonic with pulse accumulation, illustrating that increased dose does not translate monotonically into deeper structures. Values unavailable numerically were digitised from the source; lines guide the eye and source uncertainty bars are not reproduced [20]. **c**, Hatch-distance optimisation during roll-to-roll process development on aluminium at 532 nm, 250 kHz and 500 mm s⁻¹, using Λ = 14.6 µm. Mean depth, developed interfacial area (Sdr) and profile kurtosis (Pku) identify different favourable processing conditions, with the source optimum located near a hatch distance of 360 µm (24Λ); 365.5 µm is reported in the accompanying main text [21]. Together, the panels show that DLIP scale-up can be achieved through enlarged optical footprint, scanner-based dose delivery or scan-path optimisation, while the resulting productivity remains meaningful only when considered together with the realised surface morphology. Source-image components in panels a and c are adapted from Refs. [64] and [21], respectively, under CC BY 4.0. Source panels were cropped, resized and combined; the underlying scientific content was not altered.

# 9. DLIP Design Atlas

The Design Atlas converts the review's causal claim into a prospective test: identify the interacting agent, nominate the written or realised variable expected to create coupling, perturb a measurable gate independently, record the direct functional endpoint and state what result would falsify the proposed relation. This sequence is stricter than an application taxonomy because every design recommendation must expose both its transfer condition and its failure test.

The strength of the resulting prediction is uneven across domains. Optical period-depth relations and the lubricated tribology aspect-ratio window provide source-specific starting hypotheses, whereas wetting, anti-icing, biological, device and manufacturing claims remain conditional on surface state, operating regime, endpoint or system architecture. Figure 6 therefore compares the literature by inference rather than by effect magnitude: each panel asks whether the variable commonly treated as predictive remains sufficient after the relevant gate or endpoint is resolved.

The six condition-resolved contrasts in Figure 6 resolve into a common diagnostic pattern. Similar relief aspect ratios span frictional benefit, neutrality and penalty until contact regime and mechanics are specified; bacterial retention separates by material, morphology and organism; and osteoblast and photovoltaic responses remain endpoint- or device-specific. For one nominal hierarchical geometry, ice adhesion changes across four coupled icing regimes, while the electrochemical condition that maximizes capacitance does not minimize OER onset potential. These contrasts do not causally identify gates. They identify candidate effect modifiers, endpoints and unresolved covariates that prospective experiments must isolate. Geometry therefore narrows the candidate design space, but prediction begins only when the proposed gate is independently perturbed and the direct functional endpoint is measured.

Figure 6 is a source-locked stress test based on the 66 condition rows that support interpretable within-study normalization; selection was independent of effect magnitude. The contrasts are therefore diagnostic rather than prospectively validating: they identify the candidate gate, unresolved covariate, endpoint or proxy that must be challenged next.

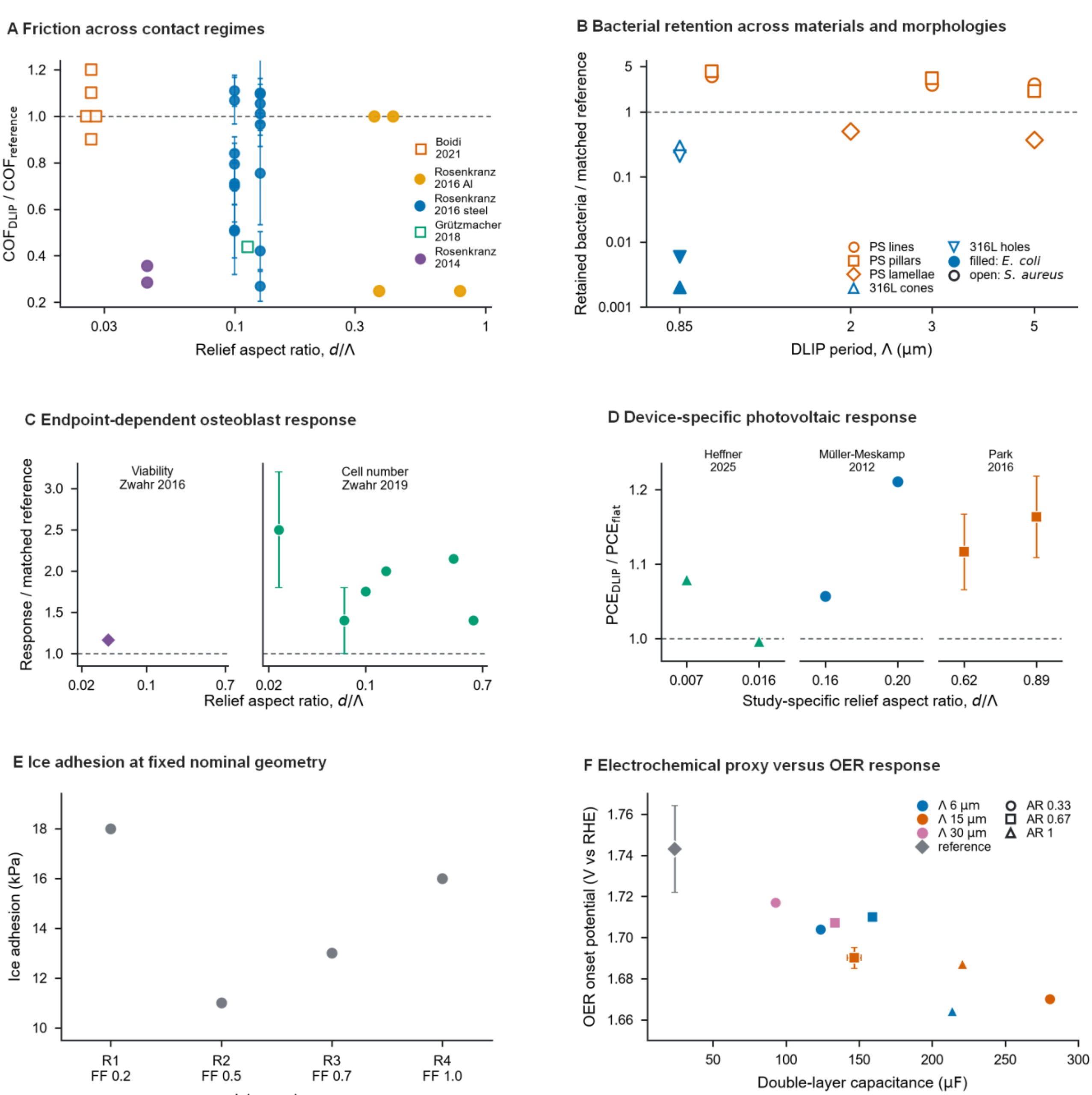

**Figure 6 | Testing the transferability of geometry–function relationships across DLIP applications.** The six panels provide condition-resolved tests of whether a geometric variable, operating state or commonly used proxy supports a transferable inference about the corresponding functional endpoint. Panels are compared by the type of inference they support rather than by effect magnitude, and numerical distances are not compared across different endpoints. a, Coefficient of friction (COF) relative to the condition-matched unstructured reference as a function of relief aspect ratio for 28 dry and lubricated conditions from five studies. Filled circles denote dry contact and open squares lubricated contact. Error bars show propagated ratio uncertainty for the Rosenkranz steel series, for which numerical uncertainties were available for both textured and reference conditions. Two coincident Boidi conditions are displaced slightly along the x-axis for visibility; their stored values are unchanged [1,16,47,49,51]. b, Retained bacteria relative to an organism-matched flat or smooth reference for 12 conditions. Orange markers represent contextual polystyrene data and blue markers DLIP-textured 316L stainless steel. Marker shape denotes morphology; filled markers represent E. coli and open markers S. aureus. Valle values were digitised from the reported log10 relative-adhesion axis and converted to linear ratios. The endpoint is bacterial retention rather than killing [3,7]. c, Human osteoblast viability and cell number relative to their respective matched references. The two endpoints are shown in separate aligned facets and are not pooled. Geometry co-varied with process-induced surface state, and error bars are shown only where numerical uncertainty was available [25,69]. d, Power-conversion efficiency relative to the matched flat device for six conditions from three photovoltaic studies. Each study is presented in a separate facet with a common y-axis and a study-specific relief-aspect-ratio axis, preserving within-study contrasts without implying a common cross-study aspect-ratio response across different texture locations and device architectures. Error bars propagate the reported patterned and planar PCE uncertainties for Park et al.; numerical uncertainty was not reconstructed for the remaining studies [54,55,70]. e, Ice adhesion strength for one nominal hierarchical Ti–6Al–4V geometry tested under four source-defined icing regimes. R1–R4 span glaze, mixed/glaze, mixed/rime and rime conditions with freezing fractions of 0.2, 0.5, 0.7 and 1.0, respectively. The regimes are plotted at equal spacing and do not define a continuous response surface or transferable optimum. Full temperature, air-speed and liquid-water-content conditions are given in the accompanying data table; the source reports error bars smaller than the markers but does not tabulate their values [2]. f, OER onset potential plotted against double-layer capacitance for the nickel-electrode series of Rox et al. Colour denotes period and marker shape the reported aspect ratio. Double-layer capacitance is used as an electrochemical-area proxy; the condition with the highest capacitance does not produce the lowest onset potential, illustrating proxy–endpoint divergence [58]. Dashed horizontal lines in a–d indicate the corresponding matched reference. Collectively, the panels show that geometry can constrain the candidate design space, while transferable prediction additionally requires resolution of contact regime, material and biological context, endpoint definition, operating state and system architecture.

**Table 2 | Prospective DLIP Design Atlas: source-bounded starting hypotheses and their falsification tests.**

| Domain | Source-bounded starting hypothesis | Candidate gate or effect modifier | Direct endpoint | Current evidence | Falsifier or present transfer limit |
|---|---|---|---|---|---|
| **Optical response** | Two source-specific examples: $\Lambda \approx 1.0$ µm and $d \approx 0.15–0.21$ µm for dot-matrix holography; $\Lambda \approx 1.74–2.59$ µm and $d \approx 0.30$ µm for structural colour [6,43] | Illumination wavelength, incidence and viewing geometry | Spectral or diffraction efficiency under stated geometry | Direct within-study optical series | Response changes at fixed relief when optical geometry changes, or depth metrology fails to predict efficiency |
| **Lubricated tribology** | $d/\Lambda \approx 0.07–0.11$ in one lubricated bearing-steel series [15] | Lubrication and contact regime; load, speed and orientation remain controlled covariates | Condition-matched COF and wear | Direct within-study series; transfer unvalidated | The window does not survive a controlled change in regime or material; no universal tribological optimum is presently supported |
| **Wetting and anti-icing** | No geometry-only range is supported; hold relief and hierarchy fixed while changing interfacial state [2,17,42] | Realised wetting or icing state, set by chemistry, ageing and environment | Dynamic wetting plus freezing, ice adhesion or de-icing energy | Direct but source-specific assays | Static contact angle improves while the relevant icing endpoint does not, or nominally fixed geometry changes sign across states |
| **Antibacterial response** | Bacteria-scale $\Lambda \approx 0.85–5$ µm and $d \approx 0.47–1.85$ µm span different materials and non-equivalent endpoints; they are not a pooled design range [3,4,7] | Material activity is a candidate primary gate; conditioning, organism and exposure are covariates | Retention and viable count reported separately | Direct retention on polymer and steel; direct killing on copper | Retention and killing diverge, or the effect disappears on an inert material at matched geometry |
| **Biomedical response** | $\Lambda \approx 3–17$ µm and $d \approx 0.4–2.1$ µm provide source-specific guidance and access hypotheses on metallic and ceramic surfaces [18,25,44] | Protein conditioning, biochemical cues and culture protocol | Alignment, migration, proliferation and phenotype reported separately | Direct coupled within-study series; phenotype transfer unvalidated | Alignment changes without the claimed phenotype, or biochemical controls reproduce the response on flat surfaces |
| **Photovoltaic and electrochemical devices** | Use within-stack relief or access metrics; no pooled cross-device optimum is supported [19,54,55] | Layer conformity, transport state, sheet resistance, electrolyte and oxide state | PCE or electrochemical current/overpotential under matched conditions | Mixed direct endpoints, models and proxies | An optical or area proxy improves while the complete-device or electrochemical endpoint does not |
| **Manufacturing translation** | Prospective target: achieved depth distribution and surface state at the stated area rate, rather than nominal rate alone [20,21,64] | Rate-dependent dose accumulation, focus, overlap, drift and post-process history | Rate-qualified morphology and the intended functional assay | Process and morphology evidence; direct function-at-rate is largely missing | Throughput increases while depth distribution, state or function moves outside the validated window |

The numerical ranges in Table 2 are source-specific starting hypotheses, not prescriptions for transfer between materials or applications. The Atlas itself is prospective: the current corpus does not causally validate its gate assignments. A candidate gate becomes operationally established only when a prospective study holds the realised geometry and relevant covariates fixed while perturbing that gate, or holds the gate fixed while varying geometry, and then measures the direct endpoint. A failed prediction is therefore an informative boundary of the framework rather than an exception to be discarded.

# 10. Benchmarks And Reporting

The Design Atlas gains predictive credibility through experiments capable of falsifying its assignments. The first priority is a fixed-geometry gate perturbation. Wetting and anti-icing studies should hold period, depth and hierarchy constant while independently varying chemistry, ageing, temperature, air speed and liquid-water content, with static wettability retained only as a screening measurement. Freezing delay, ice adhesion and de-icing energy must remain separate endpoints [2,42].

The complementary benchmark is a fixed-gate geometry series. Tribology should hold material, lubricant, load and speed constant while varying depth or aspect ratio at fixed period, then repeat the series across a deliberately changed film or contact regime. This design tests whether the reported $d/\Lambda$ window is mechanistic or source-specific and distinguishes depth-dependent effects from generic texture-benefit claims [15,16,47].

Biological and device studies require endpoint separation. Antibacterial experiments should report retention and viable count independently across material classes; biomedical experiments should separate alignment and migration from proliferation and phenotype. Photovoltaic and electrochemical studies should pair every optical, area or transport proxy condition-by-condition with PCE, current or overpotential. A proxy becomes predictive only if its relation to the direct endpoint survives a controlled change in geometry or gate [3,4,25,55,58].

Manufacturing provides the decisive transfer benchmark: reproduce the same functional window while area rate is increased. Period, depth distribution, surface state and the direct functional endpoint must be sampled at the operating rate, not in a separate morphology series. This experiment would distinguish faster pattern writing from reproducible functional manufacturing and provide the calibration target for autofocus, diffraction and emission monitoring [21,23,66].

All four benchmarks depend on a common minimum record: period; depth or d/Λ distribution; hierarchy; pulse regime and accumulated dose; material, storage, ageing and post-processing state; assay conditions; matched reference; uncertainty; and, for scale-up, processed area, throughput and sampling strategy. Negative results become reusable only when this record is complete. A failed icephobic state, increased bacterial retention, proxy-endpoint divergence or loss of function at speed then defines a quantitative transfer boundary rather than an unexplained exception [30,71].

# 11. Conclusions

Across the diverse applications examined in this review, a consistent picture emerges. Periodic geometry defines how a surface can interact with its surroundings, yet it rarely determines function on its own. Similar DLIP textures can reduce friction, delay icing, retain bacteria, promote bacterial killing, guide mammalian cells or improve optical and electrochemical performance, even when their characteristic dimensions are comparable. These apparent inconsistencies do not reflect contradictory evidence. They arise because geometry is only one part of the functional interface.

This distinction provides a different way of interpreting DLIP. The fabrication process first prescribes a written geometry through the optical interference field. Material response subsequently transforms that geometry into a realised interface defined by relief depth, hierarchy, surface chemistry and processing history. Functional performance emerges only when this realised interface couples to an interacting agent under a specific interfacial state. Written geometry, realised interface and function therefore represent successive stages of the same causal sequence rather than interchangeable descriptions of a textured surface.

Viewing DLIP through this sequence resolves several recurring ambiguities in the literature. It explains why similar geometries remain transferable in optical systems and controlled mechanical contacts, while wetting, anti-icing, antibacterial, biomedical and device-level functions additionally depend on chemistry, ageing, biological conditioning, operating conditions or system architecture. It also clarifies why apparently conflicting studies often describe different mechanisms rather than contradictory outcomes. Once the relevant interacting agent and interfacial state are identified, many inconsistencies become expected consequences of the underlying physics rather than exceptions requiring separate explanation.

The principal implication is methodological rather than technological. Progress in DLIP should no longer be evaluated primarily by the number of functional demonstrations or increasingly complex surface architectures, but by the strength of the mechanistic evidence supporting transferable design rules. A predictive relationship requires more than a correlation between morphology and performance. It requires experiments that distinguish the variables responsible for coupling from those governing its expression, identify the relevant interfacial gate through controlled perturbation, and evaluate the resulting function using direct, application-appropriate endpoints. Under this framework, negative and contradictory results become equally valuable because they define the limits within which a proposed design rule remains valid.

The same perspective extends naturally to manufacturing. Reproducible functional surfaces cannot be defined solely by processing rate or geometric fidelity. Industrial translation requires reliable control of the complete realised interface, including relief depth, hierarchy, surface state, durability and the metrology needed to verify that these properties remain within the validated functional window. Manufacturing maturity is therefore measured not only by the ability to reproduce periodic structures, but by the ability to reproduce the mechanism through which those structures generate function.

Ultimately, this review argues that the next stage of DLIP will be driven less by discovering new applications than by establishing transferable mechanistic understanding across existing ones. Predictive surface engineering begins when geometry, interface and function are treated as distinct but experimentally connected elements of the same design problem. Under this perspective, DLIP becomes more than a technique for writing periodic structures; it becomes a platform for understanding how written geometry is transformed into reproducible function..

**Data Availability**

The datasets used and/or analysed during the current study are available from the corresponding author on reasonable request.

**Acknowledgements**

*This work was co-funded by the European Union and the state budget of the Czech Republic under the project LasApp CZ.02.01.01/00/22_008/0004573.*

**Competing interests**

The author(s) declare no competing interests.